\documentclass[a4paper,fleqn,usenatbib,useAMS]{mnras}
\usepackage{color}
\usepackage{graphicx}
\usepackage{amsmath}
\usepackage[T1]{fontenc}
\usepackage{ae,aecompl}
\usepackage{graphicx}
\usepackage{amssymb}
\newcommand\sw{Swift~J0243.6+6124~}

\title[{\it NuSTAR} observation of Swift~J0243.6+6124]
{Understanding the spectral and timing behavior of a newly discovered transient X-ray pulsar Swift~J0243.6+6124}
\author[Jaisawal et al.]
{Gaurava K. Jaisawal$^{1}$\thanks{gaurava@space.dtu.dk}, Sachindra Naik$^2$ and J{\'e}r{\^o}me Chenevez$^1$\\ 
$^1$ National Space Institute, Technical University of Denmark, Elektrovej 327-328, DK-2800 Lyngby, Denmark\\ 
$^2$ Astronomy and Astrophysics Division, Physical Research Laboratory, Navrangapura, 
Ahmedabad - 380009, Gujarat, India\\}

\begin{document}

\date{}

\maketitle

\begin{abstract}

We present the results obtained from timing and spectral studies 
of the newly discovered accreting X-ray binary pulsar \sw using a 
{\it NuSTAR} observation in 2017 October at a flux level of 
$\sim$280~mCrab. Pulsations at 9.85423(5)~s were detected in the
X-ray light curves of the pulsar. Pulse profiles of the pulsar were 
found to be strongly energy dependent. A broad profile at lower 
energies was found to evolve into a double peaked profile in 
$\ge$30keV. The 3-79~keV continuum spectrum of the pulsar was 
well described with a negative and positive exponential cutoff 
or high energy cutoff power law models modified with a hot blackbody 
at $\sim$3~keV. An iron emission line was also detected at 6.4~keV 
in the source spectrum. We did not find any signature of cyclotron 
absorption line in our study. Results obtained from phase-resolved 
and time-resolved spectroscopy are discussed in the paper.

\end{abstract}

\begin{keywords}
stars: neutron -- pulsars: individual: \sw -- X-rays: stars.
\end{keywords}

\section{Introduction}

Accretion-powered X-ray pulsars are among the brightest transient 
sources in the Galaxy. They were discovered in early seventies and 
known to be powered by accretion of mass into the enormous gravitational 
field of the compact objects (\citealt{Paul2011} for a review). Most of 
these transient pulsars belong to the class of high mass X-ray binaries 
(HMXBs) in which a magnetized neutron star (B$\sim$10$^{12}$~G) co-rotates 
with a supergiant or a Be type optical companion around the common center 
of mass. The strong field lines from the neutron star interact closely 
with the accreting plasma at the magnetospheric radius and channel 
the accreted matter on to the magnetic poles. This leads to the formation 
of hot spots or column like structures at the magnetic poles of the neutron  
star which act as a source of immense high energy radiations from the pulsars 
\citep{Becker2007}.

Depending on the nature of the optical companion, HMXBs can be broadly 
classified into (i) supergiant X-ray binaries and (ii) Be/X-ray binaries 
({\citealt{Paul2011, Reig2011}). In the first case, the neutron star 
orbits a supergiant companion of O or B spectral type and accretes matter 
from its dense stellar wind, while the compact object in Be/X-ray binaries 
is associated with a non-supergiant companion of B spectral type that shows 
emission lines in their optical and infrared spectra. These emission lines 
are originated from a circumstellar decretion disk around Be stars 
\citep{Reig2011}. The neutron star in these systems accretes matter while 
passing close to or through the circumstellar disk of the companion which 
results in strong X-ray outbursts. Be/X-ray binaries show normal (Type-I) or 
giant (Type-II) outbursts in which X-ray luminosity of pulsar gets enhanced 
by a factor of 10 or more. Type-I outbursts ($\sim$10$^{37}$~erg/s) are 
periodic and coincide with the periastron passage, whereas Type-II outbursts 
($\ge$10$^{38}$~erg/s) are rare and do not show any orbital dependency 
(\citealt{Reig2011} and reference therein). The spin period of the neutron 
star in these systems ranges from a few second to 1000~s and is known 
to follow a positive correlation with orbital period in the Corbet diagram 
\citep{Corbet1986}. 

The new X-ray transient \sw was discovered by {\it Swift} observatory on 
3 October 2017  at a flux level of $\sim$80~mCrab \citep{Kennea2017}. 
The source was first thought to be a gamma ray burst (GRB) or Galactic 
transient \citep{Cenko2017}. However, the increasing flux from continuous 
monitoring ruled out the possibility of a GRB. X-ray pulsations at 
$\sim$9.86~s were detected from the source using data from {\it Swift}/XRT 
and {\it Fermi}/GBM (\citealt{Kennea2017, Jenke2017}). This confirmed the 
nature of \sw as a pulsating neutron star \citep{Bahramian2017}. Within the 
X-ray error box of the transient source, a B type star (USNO-B1.0~1514-0083050) 
was detected which is most likely  the optical counterpart of the system 
(\citealt{Kennea2017, Stanek2017, Yamanaka2017}). Further monitoring of the 
companion showed a tentative presence of hydrogen and helium emission lines 
in the optical spectra, indicating the system to be a Be/X-ray binary 
\citep{Kouroubatzakis2017}. Using optical observations of the Be 
counterpart of the X-ray source, the distance of the system is reported 
to be 2.5 \citep{Bikmaev2017} whereas data from Gamma-ray Burst Monitor 
(GBM) onboard {\it Fermi} obtained during the October 2017 X-ray outburst 
yielded the source distance of 4~kpc \citep{Doroshenko2017}. In this work, 
we have studied a detail spectral and timing properties of the pulsar for the 
first time by using a {\it Nuclear Spectroscopy Telescope Array} ({\it NuSTAR}) 
observation during 2017 October outburst. A description about observation, 
results and discussion are presented in Sections 2, 3 and 4 of the paper, 
respectively.

\section{Observation and Analysis}

Following the discovery of the transient source and its interesting X-ray activity, 
a Target of Opportunity (ToO) observation of \sw was performed with {\it NuSTAR} 
on 5 October 2017 \citep{Harrison2013}. The observation was carried out for a 
total elapsed time of  $\sim$37~ks with {\it NuSTAR} equivalent to an effective 
exposure of $\sim$14.3~ks (Obs-id: 90302319002). During the observation, the 
source was detected in the rising phase of the outburst at a flux level of 
$\sim$280~mCrab in 15-50 keV range with {\it Swift}/BAT.
 
The first hard X-ray focusing observatory, {\it NuSTAR}, was launched in June 2012 
by NASA \citep{Harrison2013}. It  covers soft to hard X-ray energy range from 3 to 
79~keV with the help of two identical grazing angle focusing telescopes. The mirrors 
of these optics are coated with Pt/SiC and W/Si multilayers allow to reflect high 
energy photons below 79~keV on CZT detectors at the two focal planes FPMA and FPMB. 
We have followed standard procedures to analyze the {\it NuSTAR} data using 
{\tt NuSTARDAS} 1.6.0 software of HEASoft (version 6.19). The unfiltered events 
were reprocessed first in our study by using {\it nupipeline} together with the updated 
version of Calibration data base (CALDB; released on 2017 October 02) files. The cleaned 
events generated after the reprocessing were used further to extract  science products 
such as light curves, spectra, response matrices, and effective area files with 
{\it nuproduct} package. The source light curves and spectra were accumulated from both 
the detectors FPMA and FPMB by considering a circular region of 180 arcsec around the 
source center. The background products were produced in a similar manner by selecting
a region away from the source.

\begin{figure}
\centering
\includegraphics[height=3.2in, width=2.2in, angle=-90]{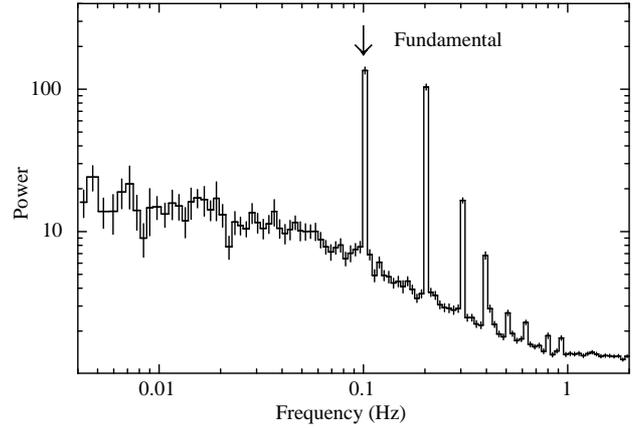}
\caption{Power density spectra (PDS) of \sw obtained from the FPMA data 
of {\it NuSTAR} observation on 5 October 2017. Pulsations at $\sim$9.85~s was 
clearly detected in PDS with its harmonics at higher frequencies. Arrow mark in 
the figure corresponds to the fundamental frequency of the pulsar in figure.}
\label{pds}
\end{figure}

\section{Results}

\subsection{Timing analysis}

Source and background light curves were extracted at a time resolution of 
100~ms by following the methods as described above. We also applied barycentric 
correction on the background-subtracted light curves to incorporate the effect 
of Earth and satellite motion relative to the barycenter of the solar system. 
The pulsation from the neutron star was searched in these light curves by using the 
{\it standard} $\chi^2$-maximization technique ({\it efsearch} task of {\tt FTOOLS}). 
We detected X-ray pulsations at 9.85423~(5)~s in the 3-79~keV barycentric corrected 
light curves from both the detectors. The distribution between the maximum 
chi-square and the trial pulsar period was fitted with a Gaussian function to 
estimate the value of pulse period and the error (1$\sigma$) associated with it. Power 
Density Spectrum (PDS; {\it powspec} tool) was also generated to confirm the 
pulsations detected in the X-ray light curves of the pulsar. In the PDS (Fig.~\ref{pds}), 
apart from the fundamental frequency corresponding to the spin period of the pulsar, 
multiple harmonics were also observed. Therefore, the pulsation at 9.85423~(5)~s was 
confirmed as the spin period of the neutron star. Pulse profiles of the pulsar 
in 3-79 keV range, were generated by folding the light curves from FPMA and FPMB
detectors at the estimated spin period and shown in the top panel of Fig.~\ref{erpp}. 
Broad-band pulse profile was found to be of complex shape. It included a minor peak 
(05--0.8 phase range) followed by the main feature at late phases (1.0--1.5 range) 
of the pulsar. To understand the energy evolution of these peaks and pulsar emission 
geometry, we have folded energy resolved light curves in different bands and presented 
in Fig.~\ref{erpp}. 

From this figure, it is clear that the pulse profiles are strongly energy 
dependent. A broad shape like profile seen at soft X-rays (3--7 keV), 
evolved into double peaked profile at hard X-rays $\ge$30~keV. Both 
the peaks were separated by approximately 180 degree phase. This suggests 
the viewing of both the poles of pulsar during observation. The energy 
evolution of the beam function can be traced successively in between low 
and high energy profiles. This evolution could also enlighten the broadening 
and phase shift in the minima of soft and hard X-ray pulse profiles (see 
Fig.~\ref{erpp}). Pulsations were detected in light curves up to 79~keV. 
From the energy evolution, it is evident that emissions from both poles 
of the pulsar were contributing and shaping the pulse profiles during 
{\it NuSTAR} observation. To quantify the fraction of X-ray photons 
contributing to the observed pulsation, we estimated pulse fraction of
the pulsar by using light curves in various energy ranges. The pulse 
fraction is defined as the ratio between the difference of maximum and 
minimum intensity to the sum of maximum and minimum intensity in the pulse 
profiles of the pulsar. Using the 3-79 keV light curve, the pulse fraction
of the pulsar was estimated to be $\sim$28.8$\pm$0.2\%. To investigate the change in 
pulse fraction with energy, we estimated pulse fraction in several energy 
ranges \& presented in Fig.~\ref{pf}. The horizontal bars on data points in the
figure represent the energy ranges in which pulse fraction has been estimated 
(see Fig.~\ref{erpp}), whereas the vertical bars represent errors on the pulse 
fraction in a given energy range. The dotted line in the figure represents
the value of pulse fraction of the pulsar in the entire 3-79 keV range. It can be 
seen that the pulse fraction of the pulsar increased towards high energy ranges, 
indicating a large fraction of high energy photons showing pulsating nature in 
the pulsar.

%
\begin{figure}
\centering
\includegraphics[height=3.1in, width=4.8in, angle=-90]{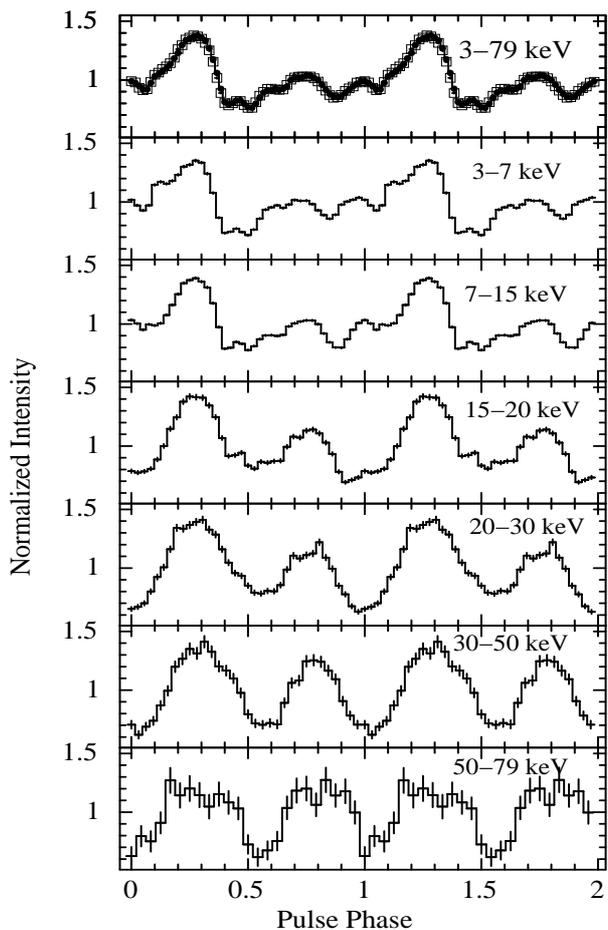}
\caption{Pulse profiles of \sw obtained from the background subtracted light 
curves of FPMA and FPMB detectors of {\it NuSTAR} observation on 5 October 2017
are shown in the top panel. The data from FPMA and FPMB are consistent with each 
other and represented by solid dot and square in the panel, respectively. Energy 
resolved pulse profiles of the pulsar from FPMA detector are also shown in the figure. 
These profiles are found to show strong energy dependence. A broad profile seen at 
soft X-rays evolved into a double peaked profile at high energies. Pulsations 
were clearly detected in the light curves up to 79 keV. The error bars 
in each panel represent 1$\sigma$ uncertainties. Two pulses in each panel 
are shown for clarity.}   
\label{erpp}
\end{figure}
\begin{figure}
\centering
\includegraphics[height=3.2in, width=1.8in, angle=-90]{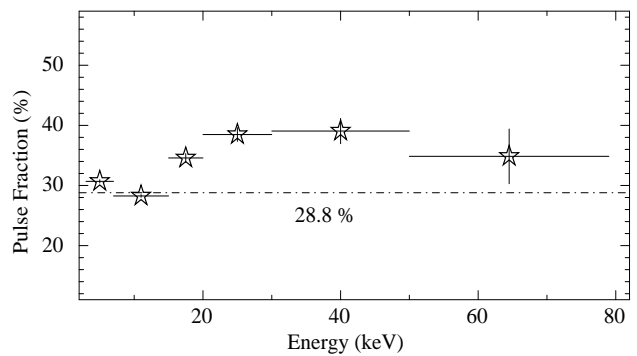}
\caption{Pulse fraction variation of the pulsar with energy. The line is 
marked at the pulse fraction obtained from the pulse profile in 3-79 keV energy range.}
\label{pf}
\end{figure}

\subsection{Pulse-phase-averaged spectroscopy}

To investigate the spectral properties of the pulsar \sw and its emission
components, phase-averaged spectroscopy was performed for the first time 
in 3-79~keV range by using data from {\it NuSTAR}. We have followed the 
procedure mention in Section~2 while extracting the spectral products. 
Source and background spectra obtained from both the detector units were 
grouped to achieve good signal to noise ratio i.e. $\ge$128 counts per 
channel bins in our study. With appropriate response, effective area and 
background files, broadband spectral study was carried out with the data 
from FPMA and FPMB detectors of {\it NuSTAR} by using {\tt XSPEC} package 
(ver. 12.9.0). All the spectral parameters for both detectors were tied 
during the fitting, while the relative normalizations of detectors were 
kept free. The cross normalization between both detectors were consistent 
with suggested value by the instrumentation team. 
   
The 3-79 keV energy spectrum of the pulsar was fitted with various 
traditional models such as power-law modified with Fermi-Dirac cutoff power 
law ({\tt fdcut}), power law model modified with high energy cutoff (HECut), 
cutoff power law ({\tt Cutoff}), Negative and Positive exponential cutoff 
power law (NPEX), and {\tt CompTT} models. Among these, NPEX is an 
empirical model which consists of two cutoff power law models with negative 
and positive photon indices. While fitting the broad-band pulsar spectrum 
with this model, the photon index of the positive power law is fixed at 2, 
whereas the negative index is kept free. This model is developed by 
\citet{Makishima1999} and widely used to describe the broad-band spectrum 
of accretion powered binary X-ray pulsars (e.g. \citealt{Terada2007, Naik2008, 
Enoto2008, Naik2011} and references therein). A component for photoelectric 
absorption was also included in the expression. All the above models failed 
to explain the pulsar continuum and produced a poor fit with reduced-$\chi^2$ 
of $\ge$2. However, addition of a blackbody component ({\tt bbodyrad} 
in {\tt XSPEC}) to these standard models, yielded acceptable fits. We 
found that NPEX, HECut and Cutoff models modified with a blackbody can 
well describe the energy spectra of the pulsar with reduced-$\chi^2$ 
close to 1 in all three cases. A fluorescence emission line from iron 
was also detected at $\sim$6.4~keV in the spectra. Any signature of 
absorption like feature or cyclotron absorption line was not seen
in the 3-79~keV continuum. Spectral parameters obtained from all three
models are given in Table~\ref{spec-para}. The value of absorption column 
density in the direction of the pulsar was found to be comparable to the 
value estimated from {\it Swift}/XRT data (\citealt{Kennea2017, 
Bahramian2017}). Based on the spectral fitting and statistics, we 
consider NPEX and HECut models with a blackbody as best fitting 
models. From the table, it can be seen that the blackbody temperature 
is relatively high $\sim$3~keV in all the cases. This suggests a complex 
continuum of the pulsar during observation in 2017 October. We have 
shown broadband energy spectrum fitted with two component (NPEX and blackbody) 
model in Fig.~\ref{sp}. The second panel in the figure indicates corresponding 
spectral residuals. We have calculated the flux using the convoluted model 
{\tt cflux} (available in {\tt XSPEC}) in this paper.

\begin{figure}
\centering
\includegraphics[height=3.2in, width=2.6in, angle=-90]{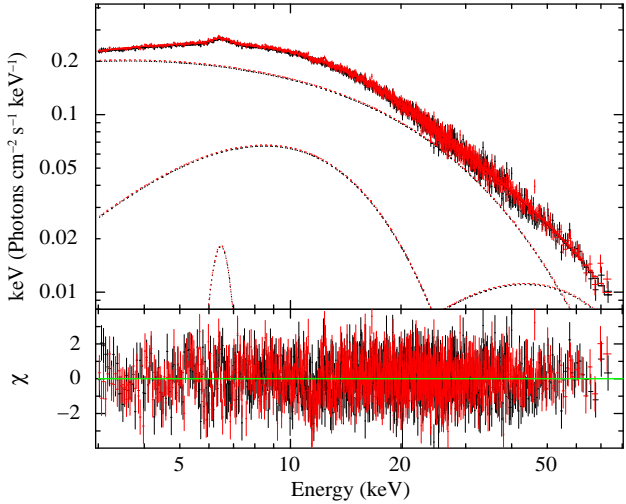}
\caption{Energy spectrum of the pulsar \sw in the 3-79~keV range obtained from 
FPMA and FPMB detectors of {\it NuSTAR} observation in 5 October 2017 along with 
one of the best-fit model comprising a NPEX continuum with blackbody and an iron 
emission line. The bottom panel shows the contribution of the residuals to the 
best fitting model.}   
\label{sp}
\end{figure}

\begin{table}
\centering
\caption{Best-fitting spectral parameters obtained from fitting 
the pulsar \sw data from a {\it NuSTAR} observation in October 2017  
with high energy cutoff power-law model, NPEX model and 
exponential cutoff power law model along with blackbody and 
a Gaussian components. The error in spectral parameters is estimated 
for 90\% confidence interval.}

\begin{tabular}{lccc}
\hline
Parameter               &HECut+BB            &NPEX+BB         &Cutoff+BB          \\
\hline
N$_{H}$$^a$             &0.9$\pm$0.3         &0.3$\pm$0.2     &0.6$\pm$0.2\\
Photon index            &1.25$\pm$0.05       &0.83$\pm$0.06   &1.10$\pm$0.03     \\
BB temp. (keV)          &3.18$\pm$0.04       &3.01$\pm$0.05   &3.04$\pm$0.03      \\
BB norm.                &1.6$\pm$0.1         &1.7$\pm$0.1     &2.04$\pm$0.06    \\
E$_{cut}$ (keV)	        &7.0$\pm$0.5         &14.5$\pm$2      &24.5$\pm$0.9        \\
E$_{fold}$ (keV)        &27.8$\pm$1.3        &--               &--    \\

\\
{\it Fe line parameters} \\
Energy (keV)            &6.41$\pm$0.04       &6.44$\pm$0.04   &6.45$\pm$0.04     \\
Width (keV)             &0.26$\pm$0.06       &0.39$\pm$0.06   &0.45$\pm$0.07       \\
Eq. width (eV)          &46$\pm$7            &71$\pm$8        &78$\pm$9            \\
\\

Flux$^b$ (3-70 keV)             &8.4$\pm$0.1       &8.4$\pm$0.1     &8.4$\pm$0.1     \\
Reduced-$\chi^2$ (dofs)         &1.06 (1573)       &1.07 (1573)     &1.11 (1574)      \\
\hline
\end{tabular}
\\
\flushleft
$^a$ : Equivalent hydrogen column density (in 10$^{22}$ atoms cm$^{-2}$); 
$^b$ : Absorption corrected flux (in 10$^{-9}$  ergs cm$^{-2}$ s$^{-1}$.) \\

\label{spec-para}
\end{table}

\subsection{Pulse-phase-resolved spectroscopy}

Phase-resolved spectroscopy was performed to understand the emission 
geometry and surrounding of the pulsar for the first time in this paper
using {\it NuSTAR} data. For this, we have accumulated the phase-sliced 
spectra in 16 phase bins using {\tt XSELECT} package. The response
matrices and effective area files as used for phase-averaged 
spectroscopy were also considered in the phase-resolved spectroscopy. 
Spectral studies were carried out in the 3-70 keV energy range for each 
phase bins of the pulsar by using source and background spectra from FPMA 
and FPMB detectors along with corresponding response and effective area 
files. The best fitting phase-averaged continuum such as NPEX and HECut 
models modified with a blackbody component were applied to describe the 
phase resolved spectra. While fitting, the width of iron line was kept 
fixed at the respective value from Table~\ref{spec-para}.  
 
Spectral parameters obtained from the phase-resolved spectroscopy are 
presented in Fig.~\ref{prs} for NPEX with blackbody continuum model. The 
first and second panels of the figure show pulse profiles in 3-10 and 
10-70 keV energy ranges, respectively. Corresponding source flux 
in these two bands are also given in bottom two panels (eighth and ninth) 
of the figure. Other spectral parameters such as blackbody temperature, 
photon index and cutoff energy are shown in third, fourth and fifth 
panels, respectively. All the parameters were varying significantly 
with pulse phase of the pulsar. Similar kind of variation in spectral 
parameters were also found with high energy cutoff model with blackbody 
in our study. Therefore, we have only presented the results from phase-resolved 
spectroscopy with composite NPEX model. From this, the equivalent hydrogen 
column density (N$_{H}$) was found nearly constant through out the pulse phases
of the pulsar. However, the value of blackbody temperature was found to be 
high (above 3~keV) in 0.15 to 0.85 phase range (see third panel in figure). 
Though, the temperature at other phases is relatively low, a bump like 
pattern was clearly evident in between 0.85 to 1.15 range of the profile. 
Corresponding to this phase range, blackbody flux was also estimated to 
be significantly high as compared to other phases of the pulsar (sixth 
panel of the figure). Moreover, photon index was getting softer as well 
in this region. The cutoff energy was found to varying in between 
10 to 30~keV with pulse phases. The power-law flux was showing a 
double peak like pattern as in hard X-ray pulse profiles. The continuum 
fluxes (absorption corrected) in 3-10 and 10-70~keV ranges were seen 
to follow the shape of pulse profiles in respective energy bands.


\begin{figure}
\centering
\includegraphics[height=2.95in, width=4.7in, angle=-90]{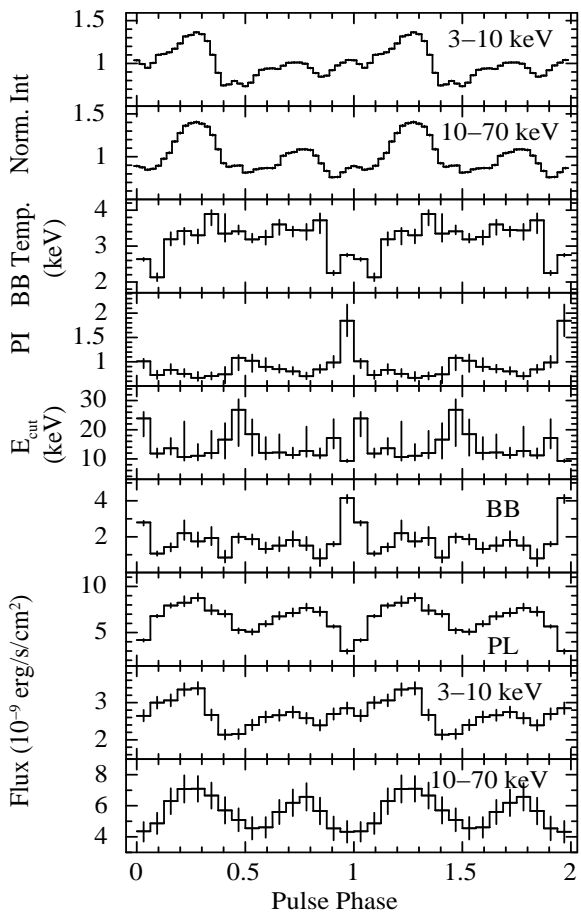}
\caption{Spectral parameters (with 90\% errors) obtained from the phase-resolved 
spectroscopy of \sw during {\it NuSTAR} observation with NPEX model along with blackbody
component. Top and second panels show the pulse profile in the 3-10 and 10-70 keV energy 
ranges, respectively. The values of blackbody temperature, photon index, cutoff energy, 
blackbody flux and power flux are shown in third, fourth, fifth, sixth and seventh 
panels, respectively. The eighth and ninth panels represent the unabsorption  
source flux in 3-10 keV and 10-70 keV energy ranges, respectively.}
\label{prs}
 \end{figure}

\begin{figure}
\centering
\includegraphics[height=2.9in, width=4.in, angle=-90]{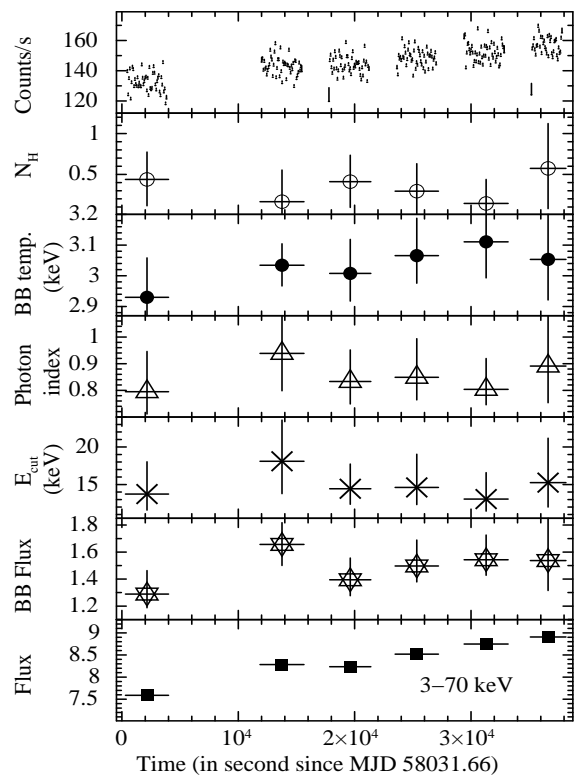}
\caption{The parameters obtained from the time resolved spectroscopy of the pulsar using
a NPEX with blackbody model during a {\it NuSTAR} observation in 2017 October. 
The first panel shows the light curve in 3-79 keV. 
Variation in column density (in unit of 10$^{22}$~cm$^{-2}$), blackbody temperature, 
photon index, cutoff energy, blackbody flux and the source flux in 3-70~keV are shown 
in second, third, fourth, fifth, sixth and seventh panels of the figure, respectively.}   
\label{time-spec}
\end{figure}


\subsection{Time-resolved spectroscopy}   

During the {\it NuSTAR} observation, the source count rate was continuously 
increasing as seen from the light curve of the pulsar (Fig.~\ref{time-spec}). 
Therefore, time resolved spectroscopy was performed to investigate 
the continuum evolution and changes in spectral parameters at different 
luminosity. Source light curve was divided in six small segments and the 
corresponding spectral products were generated by providing good time 
interval for each of them. We have used a NPEX model with blackbody 
component in the study that provided a best fit for spectra in the 
3-79 keV. The hardness ratio was also checked by dividing the light 
curve of 10-70 keV to 3-10 keV energy range. Any significant changes 
in the ratio was not detected. 

The parameters obtained from time resolved spectroscopy are shown 
in Fig.~\ref{time-spec} with the 3-79 keV light curve from FPMA detector 
of {\it NuSTAR} in the first panel. The value of column density (in unit of
10$^{22}$~cm$^{-2}$), blackbody temperature, photon index, cutoff energy, 
blackbody flux and the source flux (in 3-70~keV) are presented in 
second, third, fourth, fifth, sixth and seventh panels of the figure,
respectively. Most of these parameters were found to be moderately variable
throughout the observation. An increasing trend in blackbody temperature 
was detected with luminosity. At the same time, the photon index was appeared 
to be relatively soft. The value of source flux was found to increase
by about 17\% at end of the observation as compared to the first segment
of light curve.

\section{Discussion and Conclusions}

We have studied spectral and timing characteristics of the newly discovered 
transient \sw using data from {\it NuSTAR} in October 2017. The detection 
of strong X-ray pulsations in our study confirms this source as a pulsar. 
During the observation, the source flux was detected to be 8.4$\times$
10$^{-9}$~erg cm$^{-2}$ s$^{-1}$. Assuming a typical distance of 2.5~kpc 
\citep{Bikmaev2017}, the pulsar luminosity can be calculated to be 
$\sim$6.5$\times$10$^{36}$~erg s$^{-1}$ in the 3-70 keV energy range. 
This is a classic luminosity at which neutron stars accrete above 
sub-critical regime or close to the critical luminosity depending on 
the magnetic field (\citealt{Becker2012}). It is believed that most of 
the emission from accretion powered pulsars are due to Comptonization of
seed photons in accretion column located on the surface \citep{Becker2007}. 
For sub-critical pulsars, the bulk Comptonization of only blackbody seed 
photons leads to high energy radiations from the hot spot. A pencil beam 
pattern is expected from the pulsar in this regime. In case of bright 
(critical or super-critical) sources, thermal and bulk Comptonization 
of seed photons (blackbody, bremsstrahlung and cyclotron radiations) 
shape soft to hard X-ray emissions in a presence of radiative dominating 
shock in the column. Therefore, beam function of these pulsars can be 
complicated or as a mixture of pencil or fan geometry in critical or
super-critical luminosity regimes. 

In the present study, pulse profiles were found to be strongly energy dependent.
Most of the Be/X-ray binary pulsars show energy dependent and complex pulse 
profiles. They also include multiple absorption dips at certain phases 
(\citealt{Jaisawal2016, Epili2017} and references therein). These dips 
are found strongly dependent on energy and are thought to originate from 
absorption by matter in the form of narrow strips that are phase-locked 
to the neutron star. We have not observed any strong dip or corresponding 
increase in the column density at pulse phases from phase-resolved spectroscopy 
of  \sw. Pulse profiles in our study were found to evolve from broad peak 
like structure to double peaked profile at higher energies. The presence of
double peaked profile suggests the emissions from both the poles during 
{\it NuSTAR}  observation. Soft X-ray radiations from these hot spots 
probably originate in a form of fan beam pattern that produced 
broad peak like profile in 3-7 keV whereas the hard X-rays were emitted 
from poles of the neutron star in pencil beam pattern. This can contribute 
to viewing of both the magnetic poles of neutron star and explain the origin of 
double peaked profiles in hard X-rays.  

Despite the complex physical mechanism for pulsar emission, the energy 
spectrum of these sources can be described with simple continuum models 
such as Cutoff, HECut and NPEX models. In the case of \sw, the  
continuum from {\it NuSTAR} data was found to be described with NPEX or 
HECut model along with a hot blackbody at $\sim$3~keV. A blackbody 
component associated with a continuum model was also detected in other 
pulsars such as EXO~2030+375 \citep{Reig1999}, RX~J0440.9+4431 
\citep{Ferrigno2013}, GX~304-1 \citep{Rothschild2017} and GX~1+4 
\citep{Yoshida2017}. Assuming a distance of 2.5~kpc, the size of 
blackbody emitting region was calculated to be $\approx$325~m. This value 
is relatively small and consistent with the typical size of accretion 
column. Therefore, we expect that blackbody component is the intrinsic 
part of the spectral continuum of the pulsar which is originated from 
the accretion column. The blackbody flux is found to be $\sim$17\% of 
the source intensity.

During the {\it NuSTAR} observation, the source 
flux was found to be increasing. We have performed time resolved 
spectroscopy to explore spectral evolution and possible state changes
in the pulsar. Our results showed that the continuum was evolving 
with luminosity and showing effect of mass accretion rate. Based on this, 
a transition between sub-critical to super-critical accretion regimes 
and corresponding spectral variation or changes in emission geometry 
can be probed by future observations at different luminosity. 

Apart from time evolution, we also mapped the surrounding and spectral 
characteristics of pulsar at different pulse-phases using phase-resolved 
spectroscopy. These parameters were found to be significantly variable. 
We detected a hot blackbody around the pulsar in a wide pulse phase range. 
Though relatively lower blackbody temperature was detected at pulse phase 
close to 1, flux of the same component was found higher. Moreover, the 
photon index was also detected softer close to phase 1. This signifies 
that hard X-ray emissions from the pulsar is relatively lower at this phase. 
However, the contribution from blackbody component is clearly evident that 
shaping strong emission in the soft X-rays as seen in the 3-10 keV pulse 
profile. 

Accretion powered X-ray pulsars are expected to be highly magnetic with field
strength of the order of 10$^{12}$~G. The detection of cyclotron resonance 
scattering features can provide a direct measure on magnetic field. We have 
not found any signature of absorption like feature in the pulsar spectrum in
the present study. Future observations at higher luminosity of the pulsar 
can lead to possible discovery of a cyclotron line(s) and constrain the magnetic 
field of the pulsar.

\section*{Acknowledgments}
We thank the referee for his/her suggestions on the paper. The research 
leading to these results has received funding from the 
European Union's Seventh Framework Programme and Horizon 2020 Research 
and Innovation Programme under the Marie Sk{\l}odowska-Curie Actions 
grant no. 609405 (FP7) and 713683 (H2020; COFUNDPostdocDTU). 
The authors would like to thank all the {\it NuSTAR} team members 
for ToO observation. This research has 
made use of data obtained through HEASARC Online Service, provided by the 
NASA/GSFC, in support of NASA High Energy Astrophysics Programs. This work 
used the NuSTAR Data Analysis Software (NuSTARDAS) jointly developed 
by the ASI Science Data Center (ASDC, Italy) and the California Institute 
of Technology (USA).

\end{document}